\documentclass{elsart}
\pdfoutput=1

\usepackage[square,comma]{natbib}
\usepackage{graphicx}
\usepackage{pxfonts}
\usepackage{lineno}
\usepackage{amssymb}
\usepackage{footnote}%!!!
\usepackage{multirow}
\usepackage{varwidth}
\usepackage{array}

\journal{}

\begin{document}

\thispagestyle{empty}
\begin{Large}
\textbf{DEUTSCHES ELEKTRONEN-SYNCHROTRON}

\textbf{\large{Ein Forschungszentrum der Helmholtz-Gemeinschaft}\\}
\end{Large}

DESY 13-162

September 2013

\begin{eqnarray}
\nonumber
\end{eqnarray}
\begin{center}
\begin{Large}
\textbf{Nonlinear undulator tapering in conventional SASE regime at
baseline electron beam parameters as a way to optimize the radiation
characteristics of the European XFEL}
\end{Large}
\begin{eqnarray}
\nonumber &&\cr \nonumber
\end{eqnarray}

\begin{large}
Svitozar Serkez$^a$, Vitali Kocharyan$^a$, Evgeni Saldin$^a$, Igor
Zagorodnov$^a$ and Gianluca Geloni$^b$
\end{large}

\textsl{\\$^a$Deutsches Elektronen-Synchrotron DESY, Hamburg}
\begin{large}

\end{large}
\textsl{\\$^b$European XFEL GmbH, Hamburg}
\begin{large}

\end{large}

\begin{eqnarray}
\nonumber
\end{eqnarray}
\begin{eqnarray}
\nonumber
\end{eqnarray}
ISSN 0418-9833
\begin{eqnarray}
\nonumber
\end{eqnarray}
\begin{large}
\textbf{NOTKESTRASSE 85 - 22607 HAMBURG}
\end{large}
\end{center}
%\end{widetext}
\clearpage
\newpage
\begin{frontmatter}



\title{Nonlinear undulator tapering in conventional SASE regime at baseline electron beam parameters as a way to optimize the radiation characteristics of the European XFEL}


\author[DESY]{Svitozar Serkez \thanksref{corr},}
\thanks[corr]{Corresponding Author. E-mail address: svitozar.serkez@desy.de}
\author[DESY]{Vitali Kocharyan,}
\author[DESY]{Evgeni Saldin,}
\author[DESY]{Igor Zagorodnov,}
\author[XFEL]{and Gianluca Geloni}

\address[DESY]{Deutsches Elektronen-Synchrotron (DESY), Hamburg, Germany}
\address[XFEL]{European XFEL GmbH, Hamburg, Germany}

\begin{abstract}
We demonstrate that the output radiation characteristics of the
European XFEL sources at nominal operation point can be easily made
significantly better than what is currently reported in the TDRs of
scientific instruments and X-ray optics. In fact, the output SASE
characteristics of the baseline European XFEL have been previously
optimized assuming uniform undulators at a nominal operating point
of 5 kA peak current, without considering the potential of udulator
tapering in the SASE regime. In order to illustrate this point, we
analyze the case of an electron bunch with nominal parameters. Based
on start-to-end simulations, we demonstrate that nonlinear undulator
tapering allows one to achieve up to a tenfold increase in peak
power and photon spectral density in the conventional SASE regime,
without modification to the baseline design. The FEL code Genesis
has been extensively used for these studies. In order to increase
our confidence in simulation results, we cross-checked outcomes by
reproducing simulations in the deep nonlinear SASE regime with
tapered undulator using the code ALICE.
\end{abstract}

%\begin{keyword}
%%edge radiation \sep near-field \sep electron-bunch diagnostics
%%\sep x-ray free-electron laser (XFEL)
%%\PACS 41.60.Cr \sep 42.25.-p \sep 41.75.-Ht
%\end{keyword}
%
\end{frontmatter}

% main text

%\linenumbers

\section{\label{sec:intro} Introduction}

The technical note \cite{TSCH} provides an overview of the design
considerations and the general layout of the X-ray instrumentation
of the European XFEL sources, beam transport systems and
instruments. Baseline parameters for the electron beam have been
defined and presented in \cite{IGO1}. These parameters have been
used for simulating FEL radiation characteristics and saturation
lengths relevant to the European XFEL SASE undulators \cite{SY1}.
The definition of saturation point used throughout \cite{TSCH} is
reported here: "Saturation is reached at the magnetic length at
which the FEL radiation attains maximum brilliance. Beyond the
saturation point, the FEL operates in an over saturated mode where
more energy can be extracted from the electron beam at the expense
of FEL parameters, including bandwidth, coherence time, and degree
of transverse coherence".

An approach based on the exploitation of the baseline electron beam
characteristics in \cite{IGO1}, together with the definition of the
saturation point reported above and with the notion that the best
FEL parameters are found at saturation has been quite useful as a
convenient starting point for the analysis of XFEL sources, beam
transport systems and instruments. However, based on recent
advancement in the FEL field, here we argue that there are two main
reasons why such approach should be modified.

First, the above-mentioned approach is based on the baseline
parameters for the electron beam \cite{IGO1} assuming an operation
point at 5 kA peak current. This choice is subjective. There is a
possibility to go well beyond the nominal peak current level. For
example, in order to illustrate the potential of the European XFEL
accelerator complex, in \cite{IGO2} considerations were focused on
an electron bunch with 0.25 nC charge, compressed up to a peak
current of about 50 kA.  An advantage of operating at such peak
current is the increase of the X-ray output peak power without of
any modification to the baseline design \cite{10TW}. The price for
using a very high peak current is a large energy chirp within the
electron bunch, yielding in turn a large (about $1 \%$) SASE
radiation bandwidth. However, there are very important applications
like bio-imaging, where such extra-pink X-ray beam is sufficiently
monochromatic to be used as a source for experiments without further
monochromatization. The example presented in \cite{10TW}
demonstrates that there is no universal choice of baseline electron
beam parameters: such choice should be considered, at least, as
dependent on the instruments.

Second, the analysis presented in \cite{TSCH} is based on the use of
uniform undulators only. A fundamental example showing the
limitations of such model for the analysis of the European XFEL is
constituted by undulator tapering effects. In fact, one obvious way
to enhance the SASE efficiency is by properly configuring undulators
with variable gap \cite{TAP1}-\cite{LAST}.

Abandoning the assumption of uniform undulators alone has far
reaching consequences. Here we illustrate the potential of undulator
tapering in the SASE regime by still considering an electron bunch
with the usual baseline parameters \cite{IGO1}, i.e. the same
assumed in \cite{TSCH}, and in all the subsequent TDR. A significant
increase in power is achievable by nonlinear rather than linear
undulator tapering \cite{LAST}. In particular, in this article we
will demonstrate that nonlinear undulator tapering allows one to
achieve up to a tenfold increase not only in peak power but also in
photon spectral density of the output radiation pulses. In order to
illustrate this point we will make an analysis for the baseline (21
cells) SASE3 undulator at the nominal electron beam energy 14 GeV.
We first optimize our setup based on start-to-end simulations for
the electron beam with 0.1 nC, compressed up to 5 kA peak current
\cite{IGO1}. In this way, the SASE saturation power in the uniform
undulator could be as large as 60 GW.  Subsequently, in order to
generate high-power X-ray pulses we exploit undulator tapering.
Tapering consists in a slow reduction of the field strength of the
undulator in order to preserve the resonance wavelength, while the
kinetic energy of the electrons decreases due to the FEL process.
The undulator taper can be simply implemented as discrete steps from
one undulator segment to the next, by changing the undulator gap. In
this way, the output power of the SASE3 undulator could be increased
from the value of 60 GW in the SASE saturation regime  to about 750
GW.  One might be surprised that the photon spectral density
increases of about a factor ten as well (see section \ref{sec:study}
for more details).

The analysis of the nonlinear FEL process refers to a problem that
can be solved only numerically. The Genesis code \cite{GENE} has
been extensively used for our FEL studies. However, an accurate
simulation of the deep nonlinear SASE regime in a tapered undulator
remains a challenging problem for numerical analysis.  During the
last decade, several additional FEL codes have been developed around
the world. In order to increase our confidence in the simulation
results, we cross-checked them with one of these, the code ALICE
\cite{ALIC}.

Summing up, in this article we demonstrate that the  performance of
European XFEL sources can be significantly improved without
additional hardware. The optimization procedure simply consists in
the optimization of the undulator gap configuration for each X-ray
beamline. Based on these findings, we suggest that new baseline
radiation parameters be defined, and that the requirements for the
instruments and beam transport systems be updated.

\section{\label{sec:study} FEL studies}

As described in the Introduction, in this paper we focus our
attention on the tremendous increase in SASE efficiency that can be
achieved using a tapering undulator technique. It must be stressed
that in actual studies we let aside our remark concerning the
optimization of the electron beam nominal working point, and we
proceed in the analysis of the benefits deriving from the
application of tapering for the baseline (21 cells) SASE3 undulator
at the nominal electron beam energy of 14 GeV, considering a 0.1 nC
bunch, compressed up to 5 kA peak current \cite{IGO1}. The main
electron and undulator parameters for simulations are shown in Table
 \ref{tt1}.

\begin{table}
\caption{Parameters for the mode of operation at the European XFEL
used in this paper.}

\begin{center}
\begin{small}\begin{tabular}{ l c c}
\hline & ~ Units &  ~ \\ \hline
Undulator period      & mm                  & 68     \\
Periods per cell      & -                   & 73   \\
Total number of cells & -                   & 21    \\
Intersection length   & m                   & 1.1   \\
Energy                & GeV                 & 14 \\
Charge                & nC                  & 0.1\\
\hline
\end{tabular}\end{small}
\end{center}
\label{tt1}
\end{table}

\begin{figure}[tb]
\includegraphics[width=0.5\textwidth]{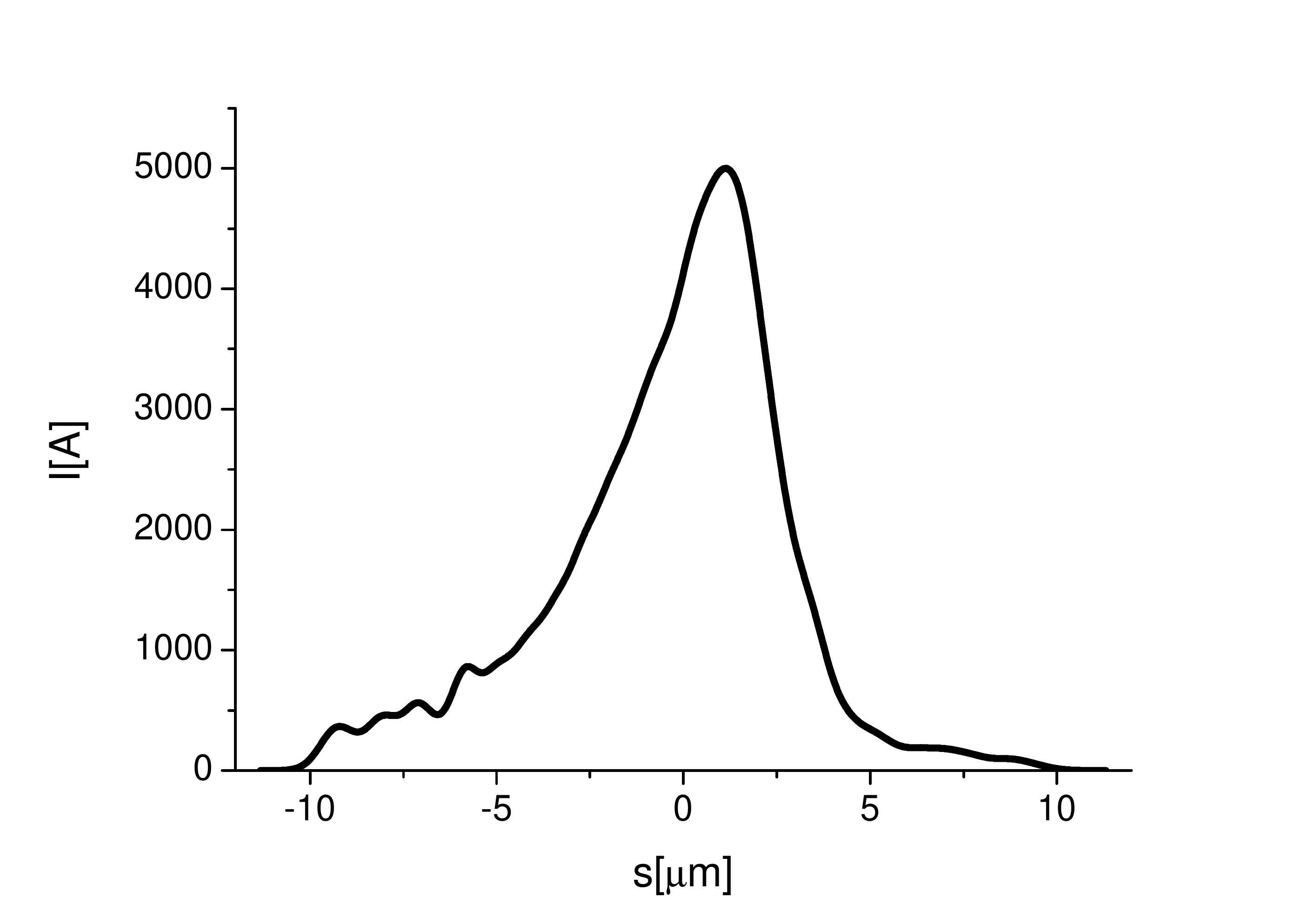}
\includegraphics[width=0.5\textwidth]{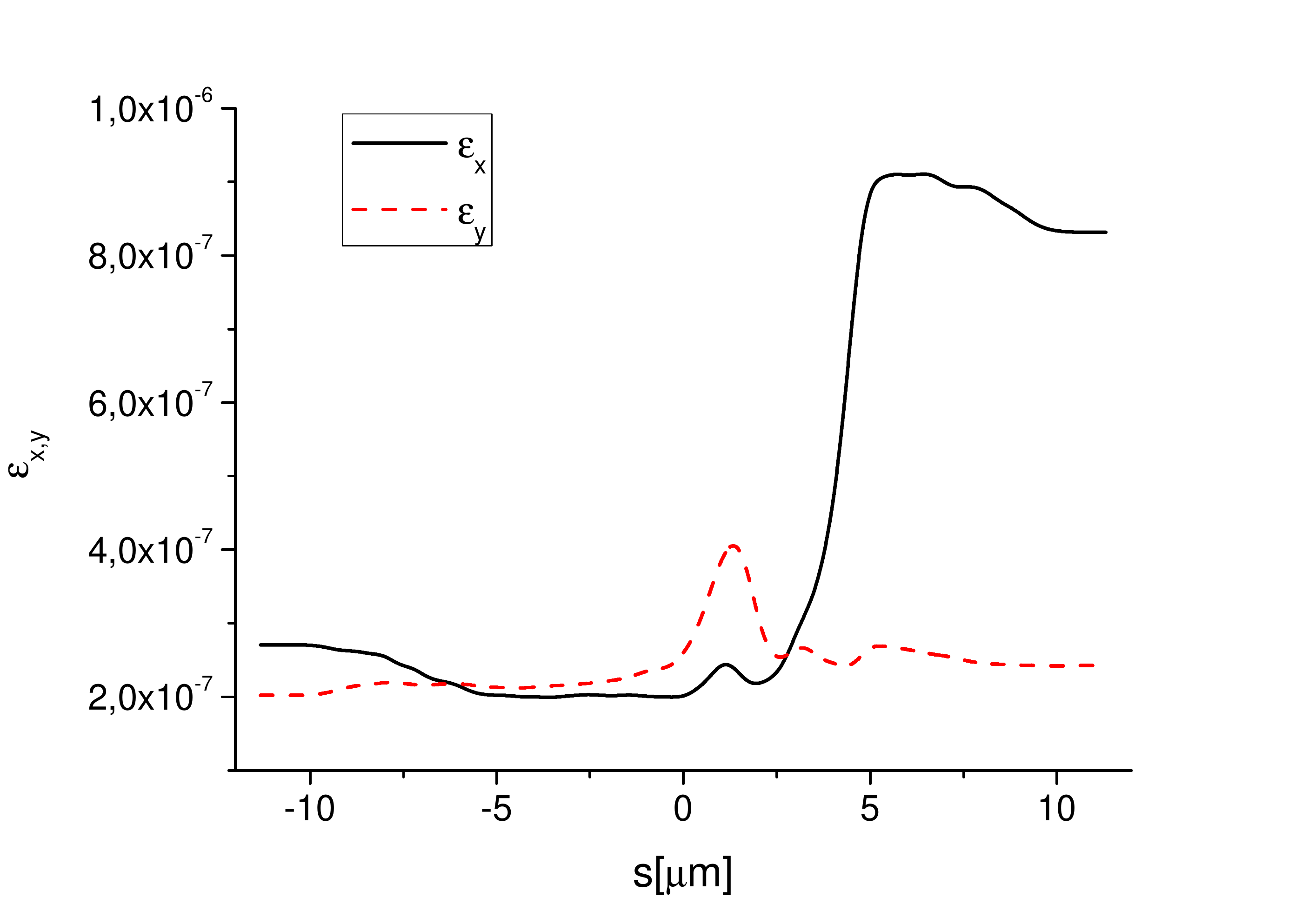}
\includegraphics[width=0.5\textwidth]{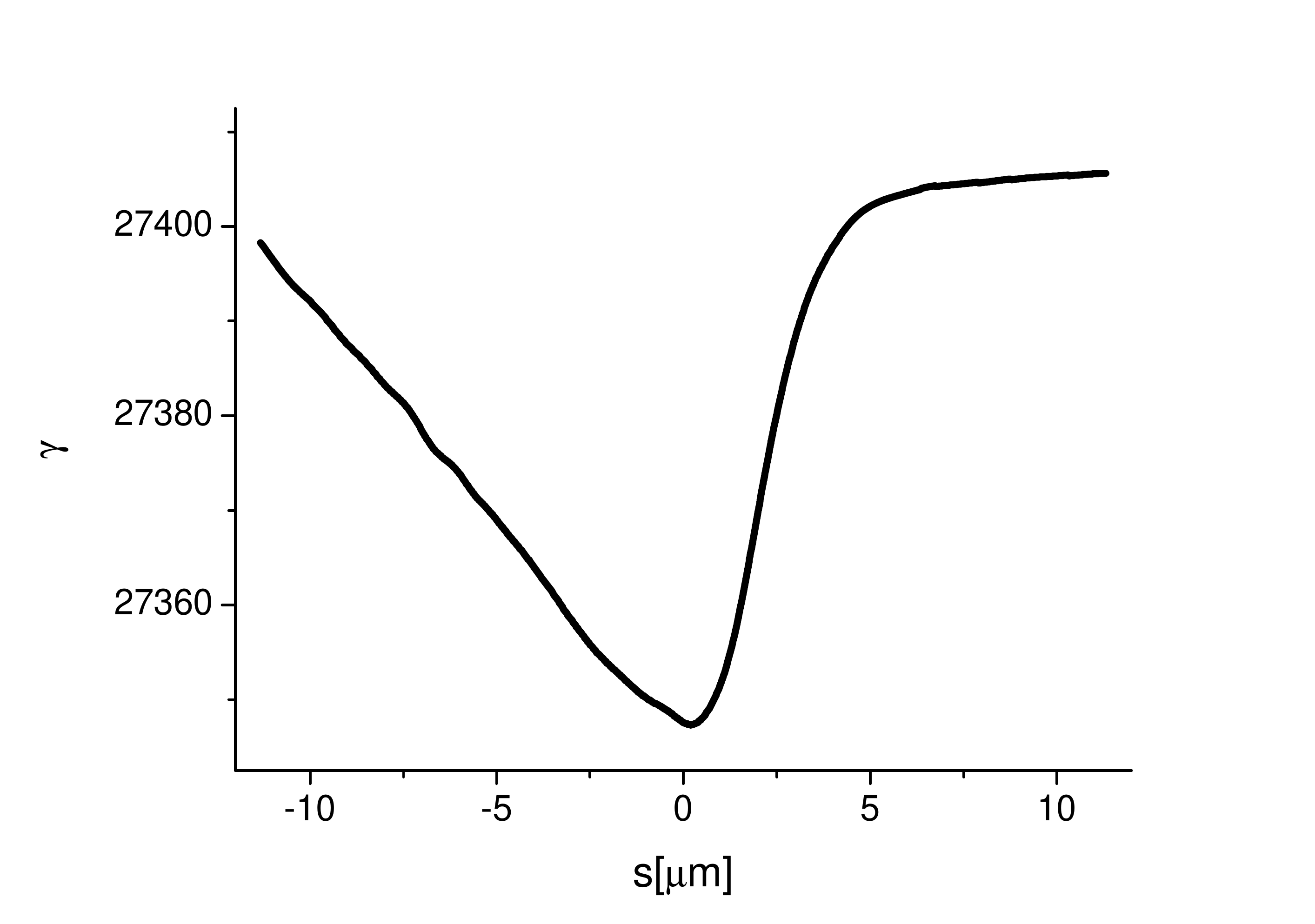}
\includegraphics[width=0.5\textwidth]{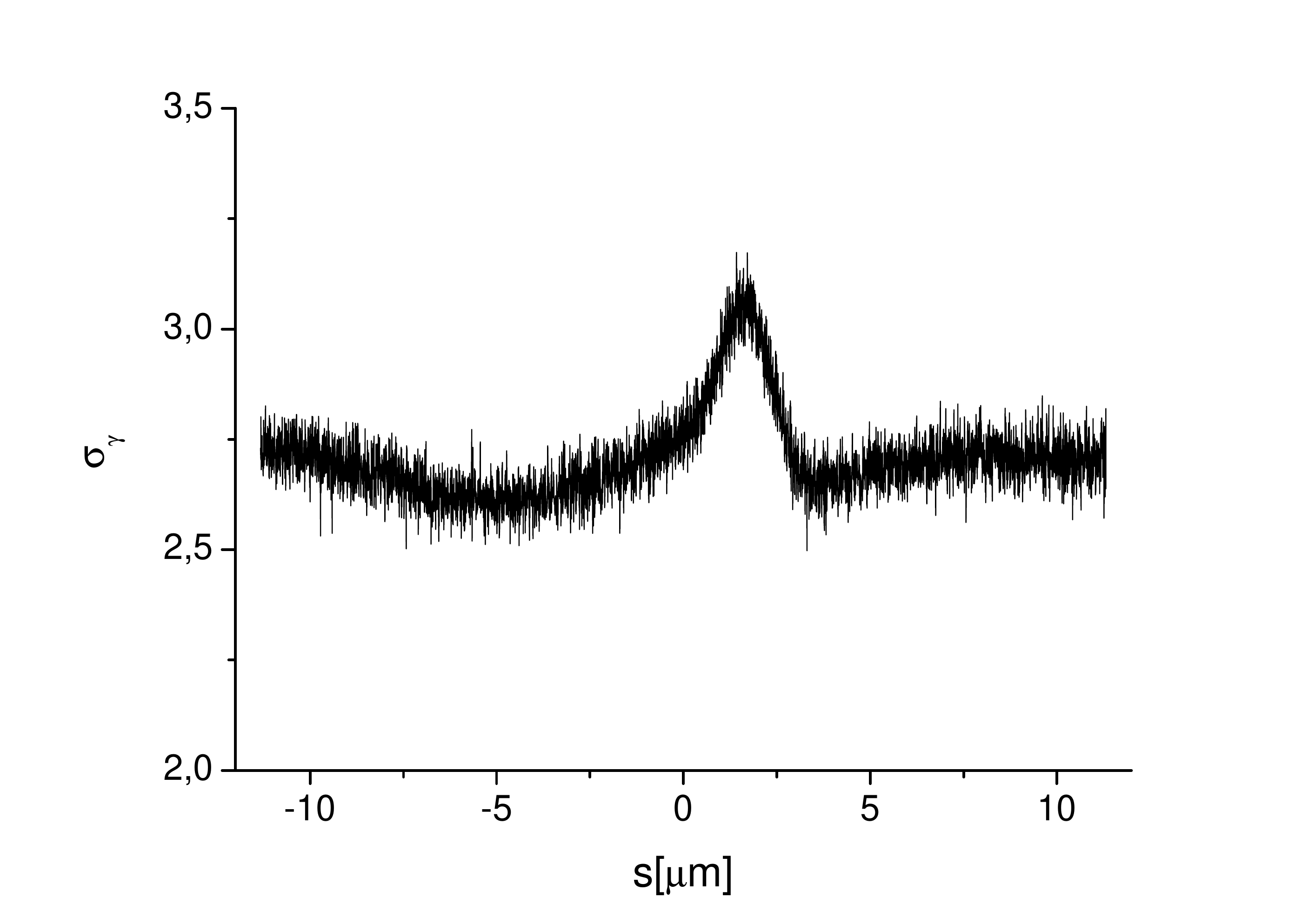}
\begin{center}
\includegraphics[width=0.5\textwidth]{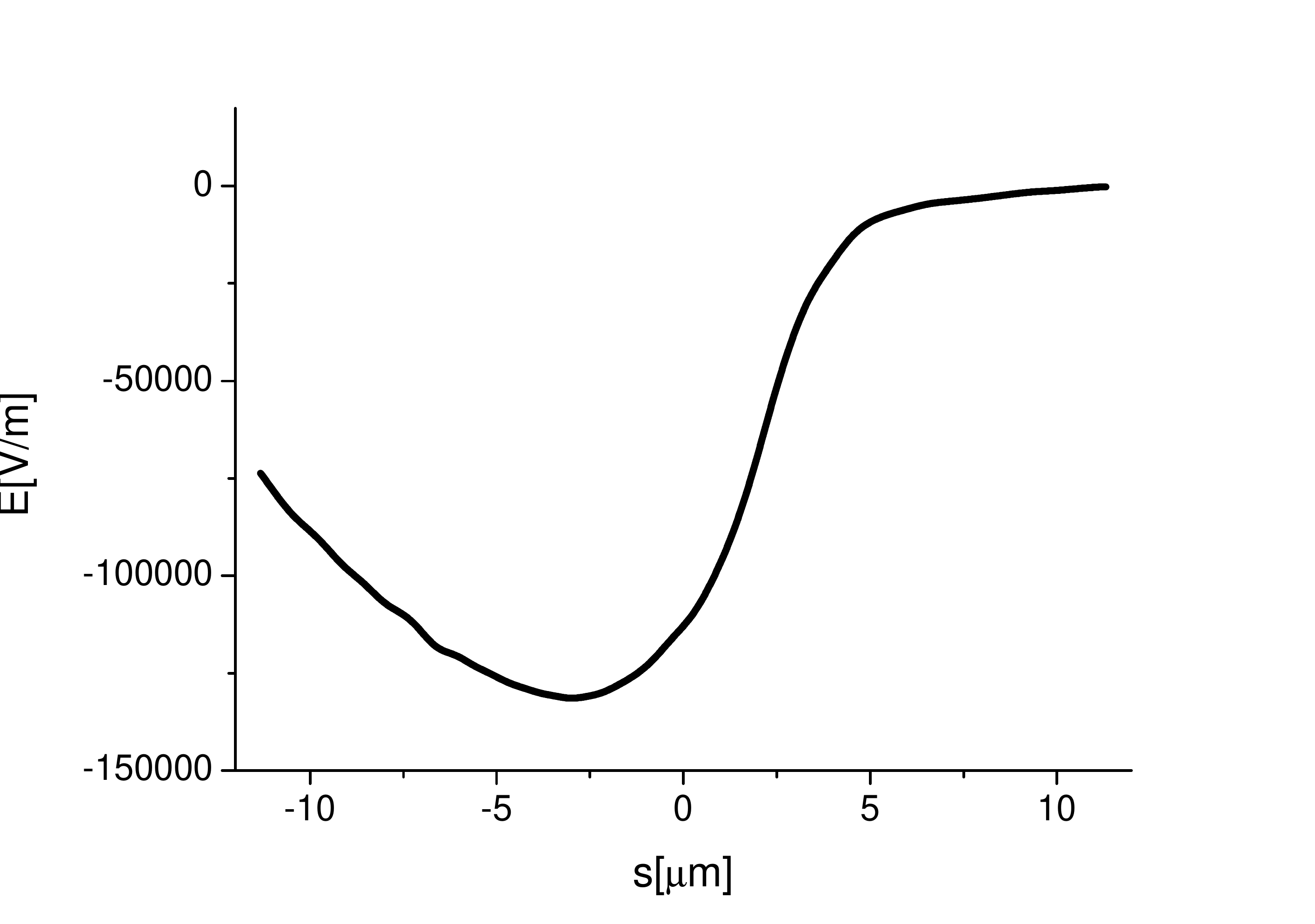}
\end{center}
\caption{Results from electron beam start-to-end simulations at the
entrance of SASE3. (First Row, Left) Current profile. (First Row,
Right) Normalized emittance as a function of the position inside the
electron beam. (Second Row, Left) Energy profile along the beam.
(Second Row, Right) Electron beam energy spread profile. (Bottom
row) Resistive wakefields in the SASE3 undulator.} \label{s2E}
\end{figure}
The nominal electron beam characteristics resulting from
start-to-end simulations are shown in Fig. \ref{s2E} in terms of
current, emittance, energy spread and energy. In addition, Fig.
\ref{s2E} also shows the resistive wake in the SASE3 undulator.
Additional energy chirp introduced by resistive wakes in the SASE1
undulator vacuum chamber are included in our simulations, as well as
quantum diffusion effects.

In this Section we show how proper undulator tapering allows one to
increase the peak radiation power of the European XFEL SASE sources
from the 100 GW power-level up to the TW power-level. In order to
perform the undulator tapering optimization we followed the
theoretical work done at DESY \cite{OURY3} and SLAC \cite{LAST},
which was used for self-seeded tapered XFELs, showing that the
baseline undulator greatly benefit of this treatment as well. To be
specific, we used the taper profile in \cite{LAST}

\begin{eqnarray}
&&K(z) = K(z_0) ~ ,~~~~~~~~~~~~~~~~~~~~~~~~~~~~~~~~~~~~
\mathrm{when} ~0 < z < z_0 ~,\cr && K(z) = [K(z_0) + d] \cdot [1 - a
(z-z_0)^b] ~, \mathrm{when}  ~ z_0 < z < L_w ~, \label{taper}
\end{eqnarray}
\begin{figure}
\begin{center}
\includegraphics[width=0.50\textwidth]{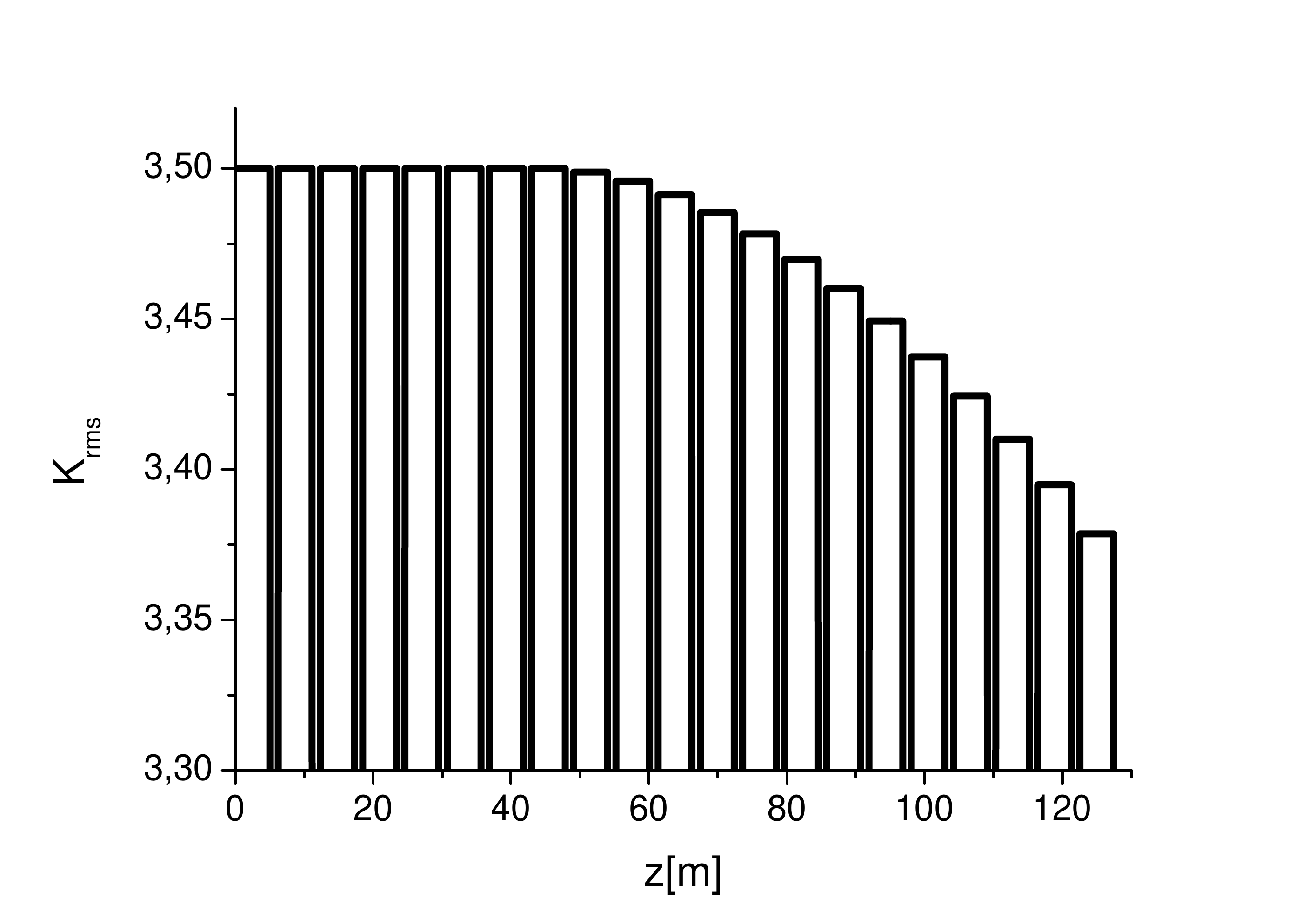}
\end{center}
\caption{Taper configuration for high-power mode of operation at
$0.6$ nm.} \label{Taplaw}
\end{figure}
\begin{figure}[tb]
\includegraphics[width=0.5\textwidth]{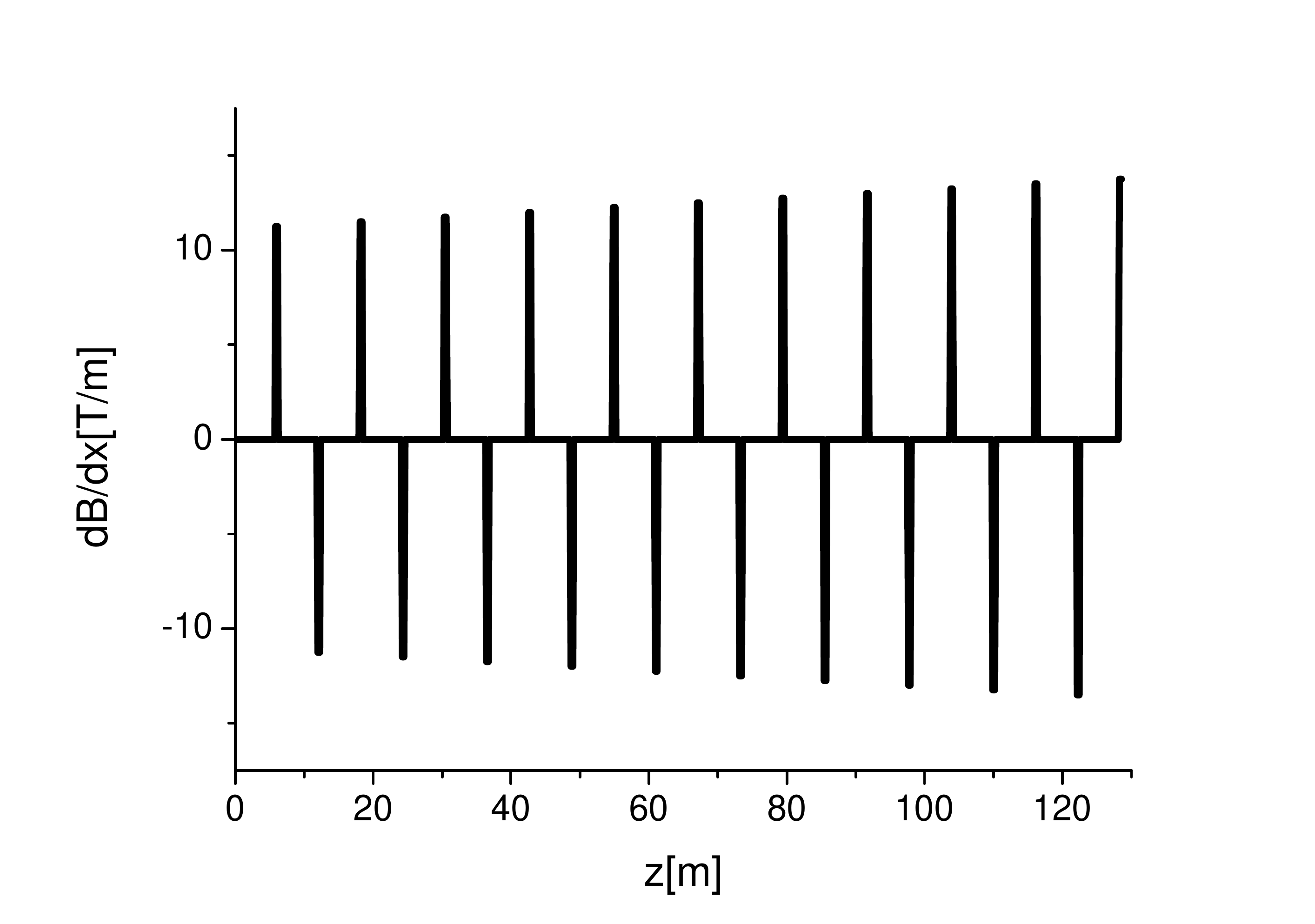}
\includegraphics[width=0.5\textwidth]{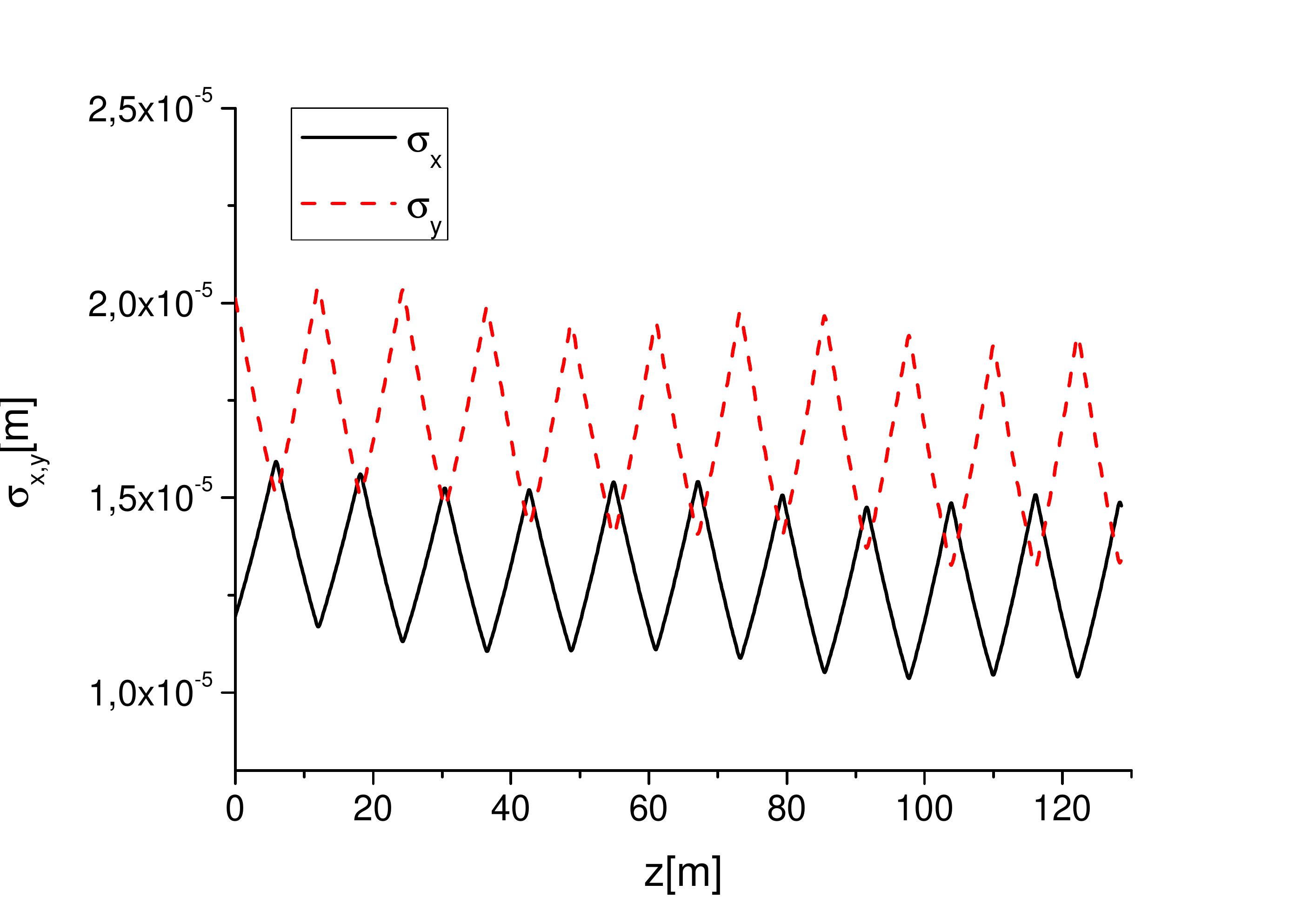}
\caption{Left plot: linear increase in the magnetic field quadrupole
gradient assumed in this paper. Left plot: consequent evolution of
the horizontal and vertical dimensions of the electron bunch as a
function of the distance inside the SASE3 undulator. The plot refers
to the longitudinal position inside the bunch corresponding to the
maximum current value. } \label{sigma}
\end{figure}
where K is undulator parameter, $z_0$ indicates the taper starting
point, $b$ is the taper profile order, $a$ is a scale coefficient,
and $d$ is the change in undulator parameter at the  tapering start
point location. Empirically, the best taper starting point was found
to be located slightly before the initial saturation point, and the
taper profile order is around $b \simeq 1.5$. This specification for
the functional dependencies of the undulator parameter allowed us to
obtained the maximum radiation power for the tapered FEL with
baseline electron beam described above by performing
multidimensional scans with Genesis SASE simulations over the
following  four parameters $z_0$, $a$, $b$, and $d$. Such
optimization resulted in the tapering law graphically shown in Fig.
\ref{Taplaw}, for radiation emitted at $2$ keV. Additionally, as
originally proposed in \cite{LAST}, we assumed a linear change,
along the undulator longitudinal coordinate $z$, in the strength of
the quadrupole field, Fig. \ref{sigma} (left) leading to the
evolution of the electron bunch transverse dimensions shown in Fig.
\ref{sigma} (right).

All simulations were performed using the code Genesis 1.3
\cite{GENE} running on a parallel machine. Results are presented for
the SASE3 FEL line of the European XFEL, based on a statistical
analysis consisting of $100$ runs, and comparing the SASE regime at
saturation with the tapered SASE regime.

\begin{figure}
\includegraphics[width=0.5\textwidth]{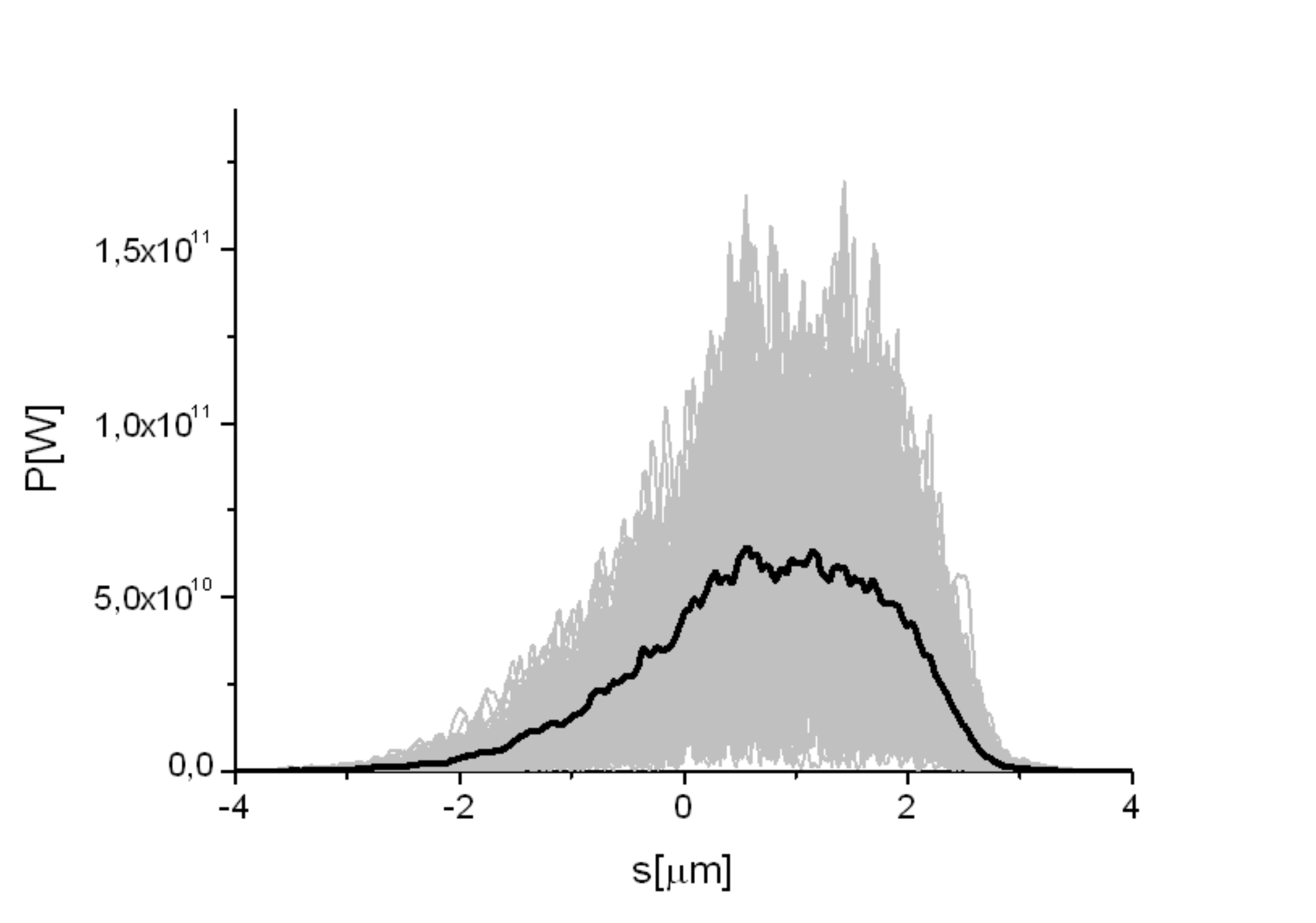}
\includegraphics[width=0.5\textwidth]{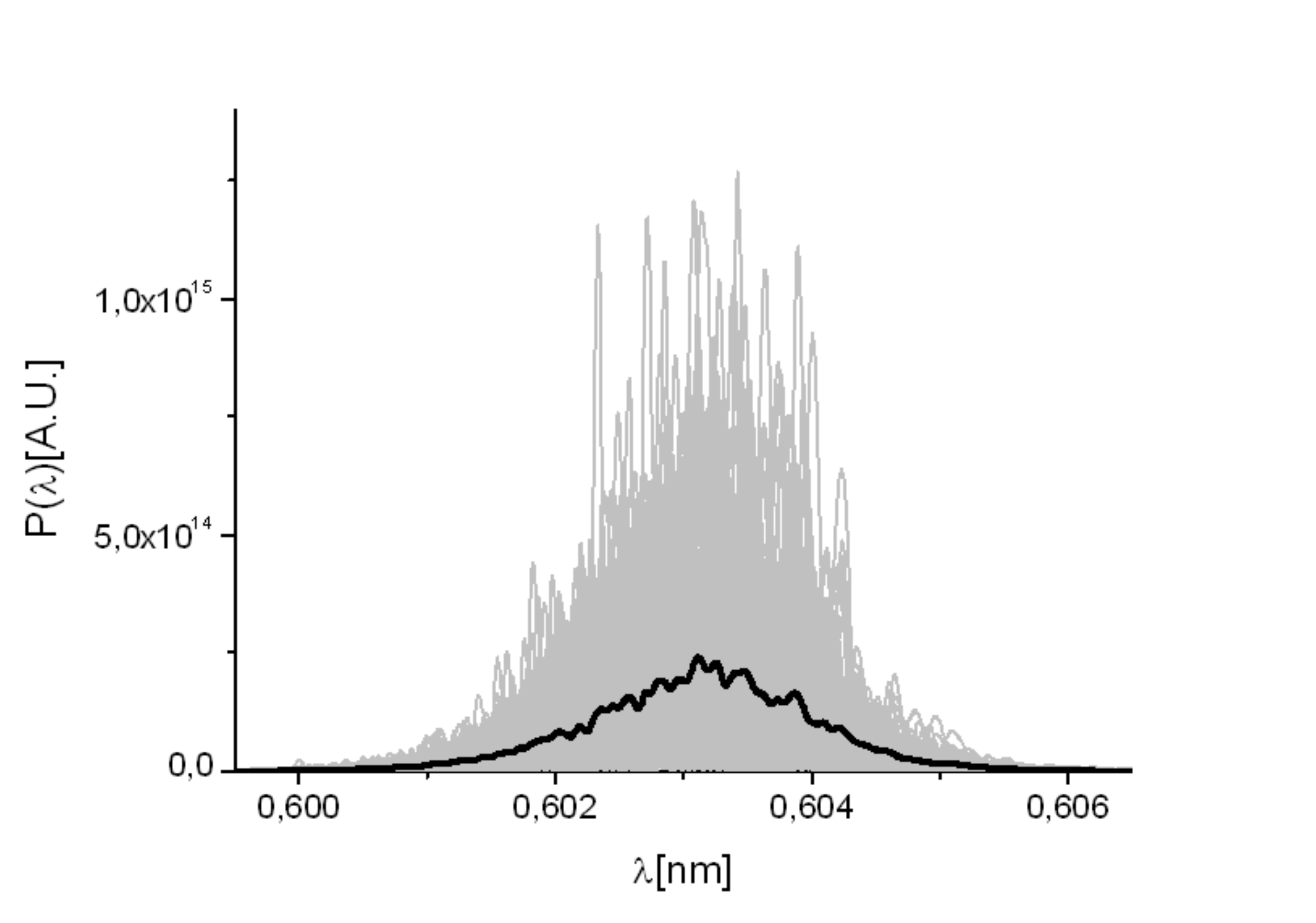}
\includegraphics[width=0.5\textwidth]{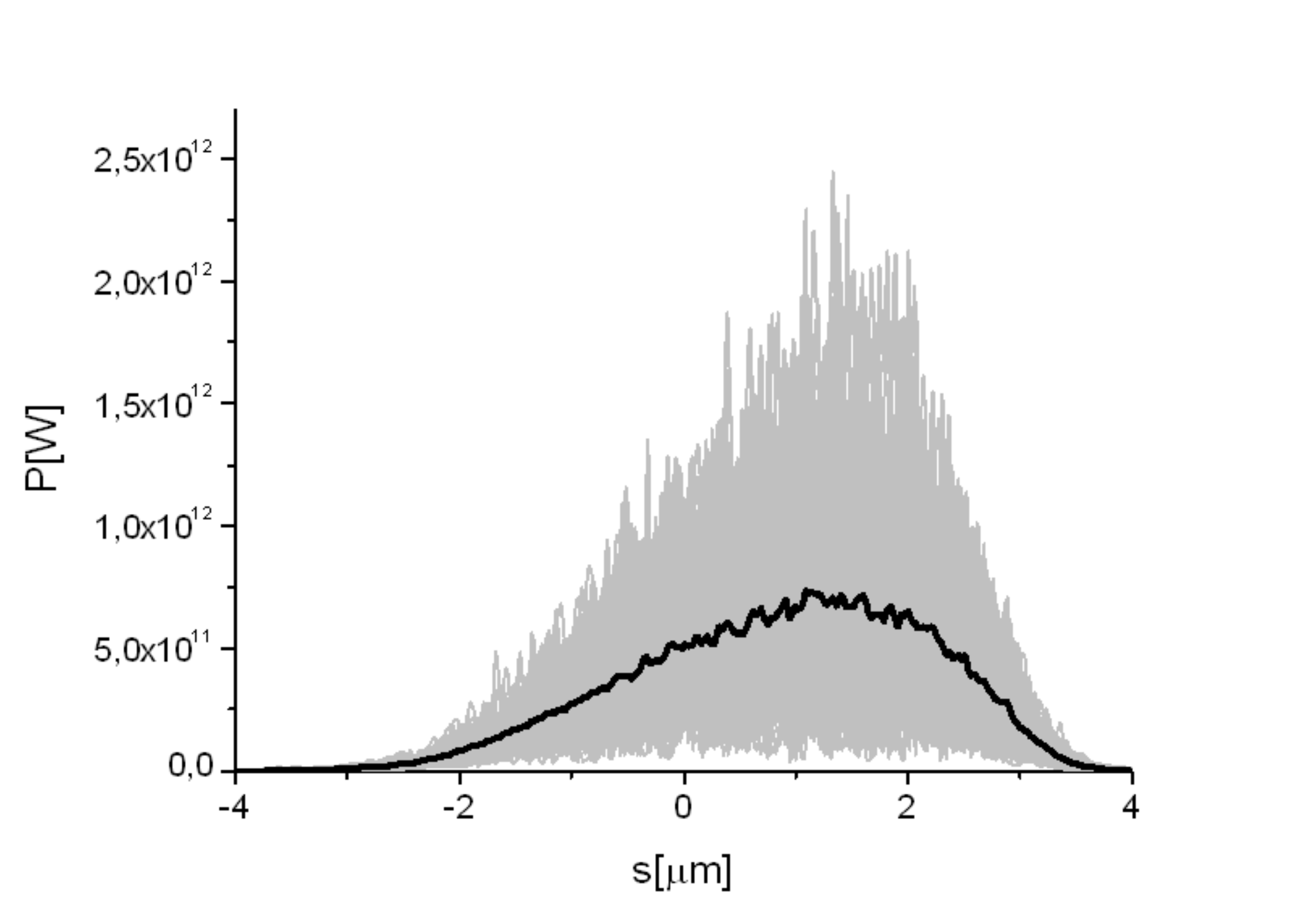}
\includegraphics[width=0.5\textwidth]{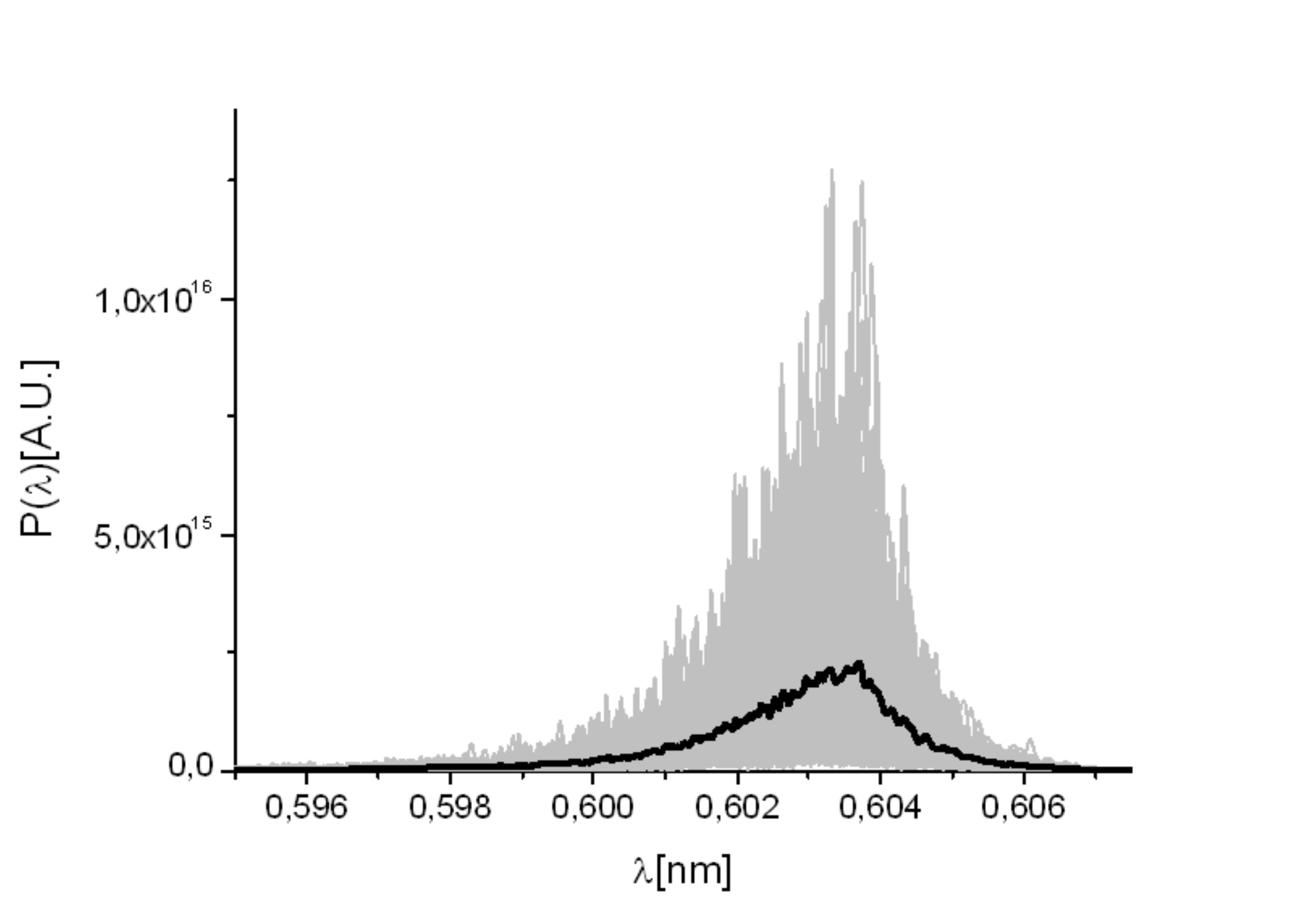}
\caption{Top row: Power and spectrum produced in the standard SASE
mode at saturation (and, therefore, without tapering). Bottom row:
Power and spectrum produced in the standard SASE mode including
post-saturation tapering. Scales of spectral energy density are the
same for both cases. Grey lines refer to single shot realizations,
the black line refers to the average over a hundred realizations. A
tenfold increase in the shot-to-shot averaged spectral energy
density can be seen by inspection.} \label{PowSpec}
\end{figure}

\begin{figure}
\includegraphics[width=0.50\textwidth]{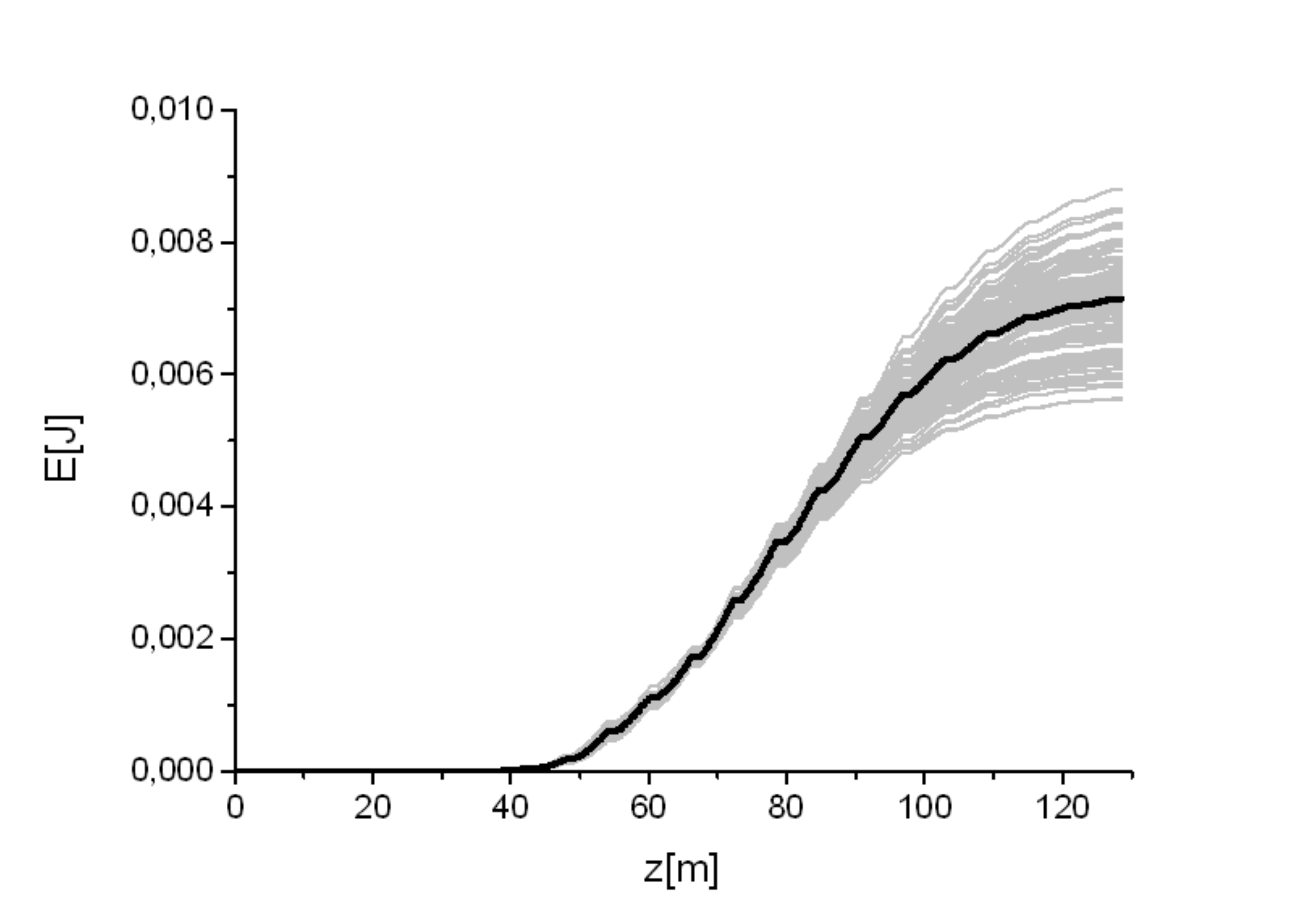}
\includegraphics[width=0.50\textwidth]{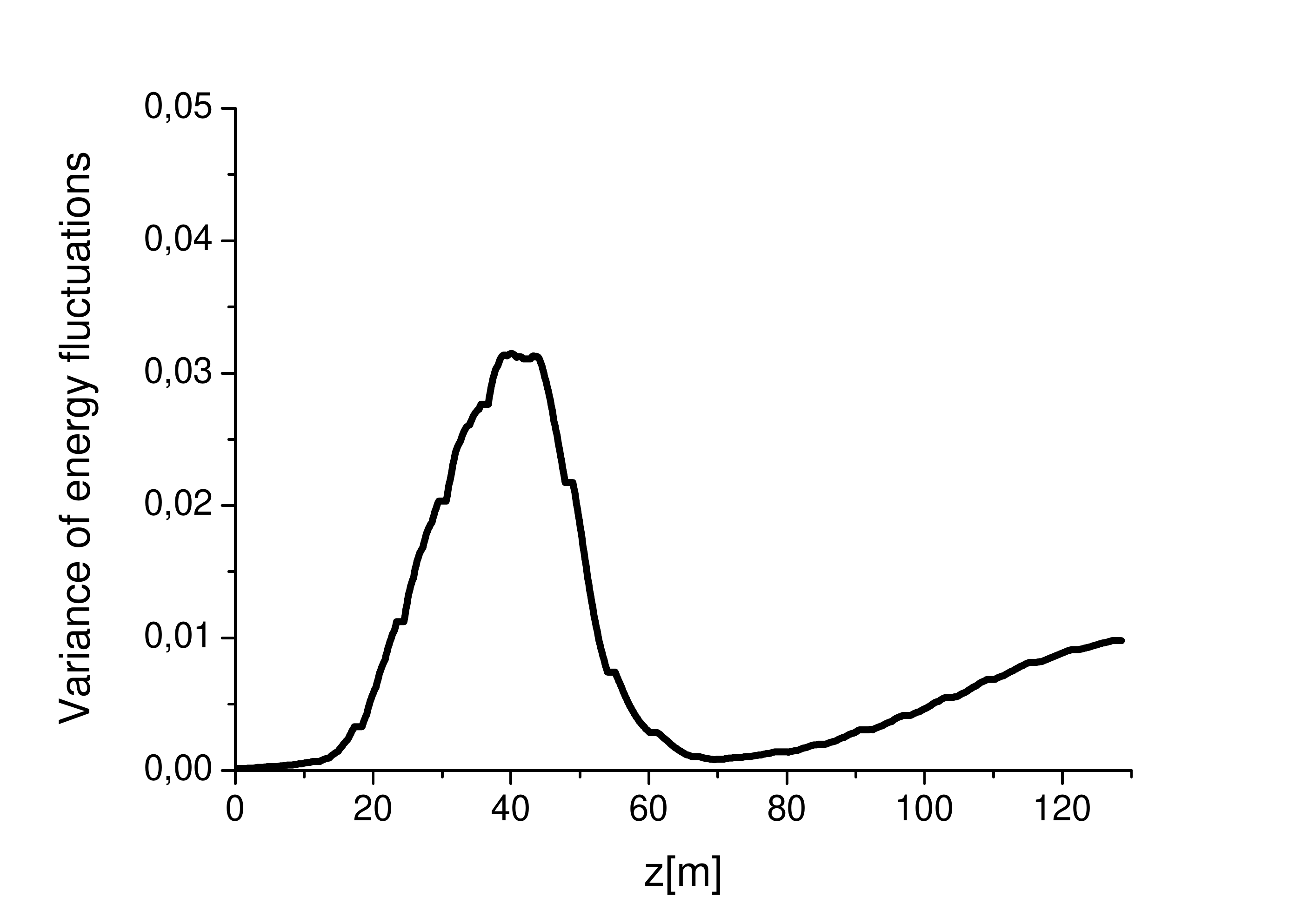}
\caption{Evolution of the output energy in the photon pulse and of
the variance of the energy fluctuation as a function of the distance
inside the output undulator, with tapering. Grey lines refer to
single shot realizations, the black line refers to the average over
a hundred realizations.} \label{EnvaroutS}
\end{figure}
Fig. \ref{PowSpec} shows a comparison of power and spectrum produced
in the standard SASE mode at saturation (and, therefore, without
tapering) and power and spectrum produced in the standard SASE mode
including post-saturation tapering. What is important to notice
here, is that a tenfold increase in the shot-to-shot averaged
spectral energy density can be seen by inspection. Fig.
\ref{EnvaroutS} shows the evolution of the output energy in the
photon pulse and of the variance of the energy fluctuation as a
function of the distance inside the output undulator, including
tapering.

\begin{figure}[tb]
\includegraphics[width=0.5\textwidth]{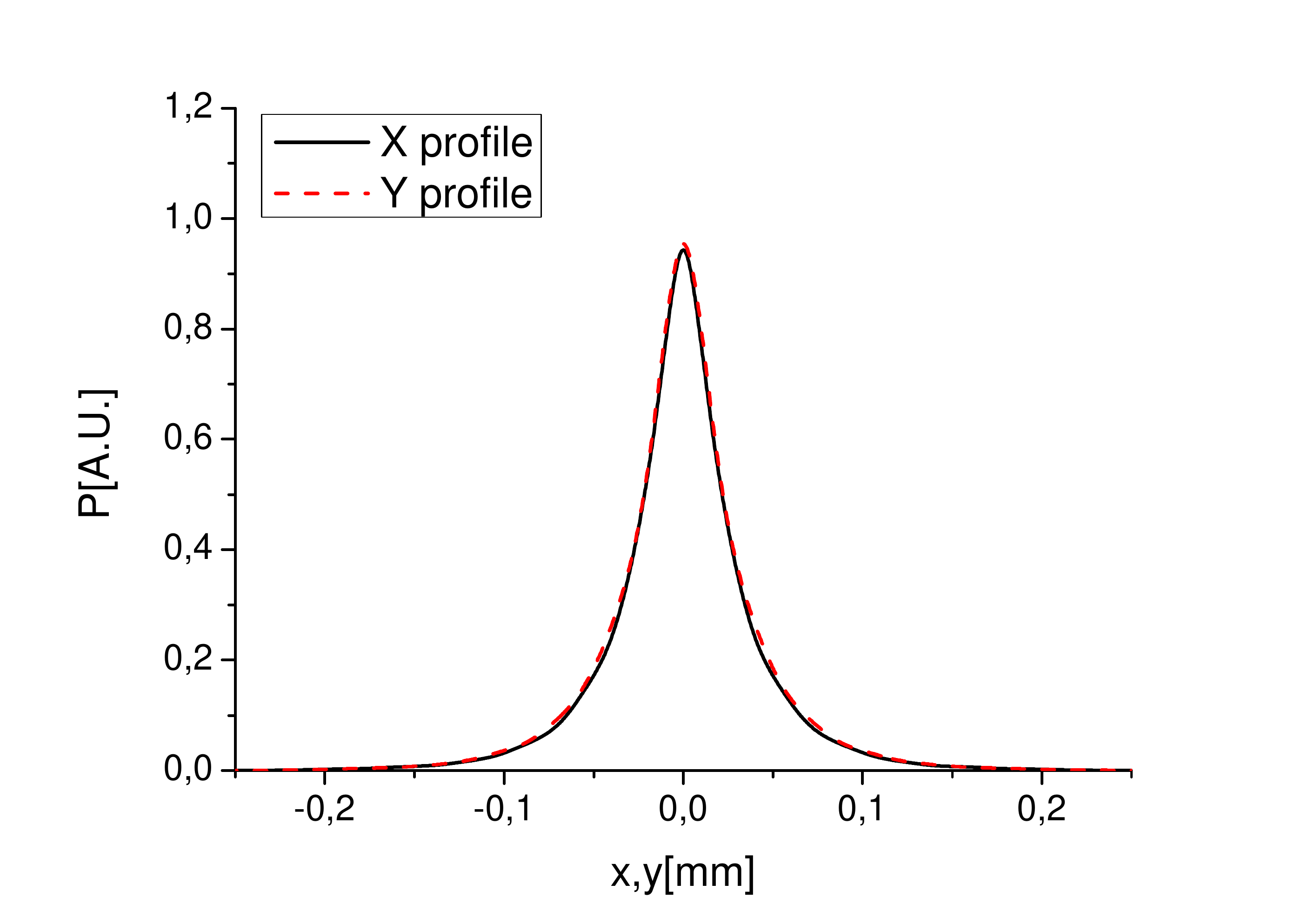}
\includegraphics[width=0.5\textwidth]{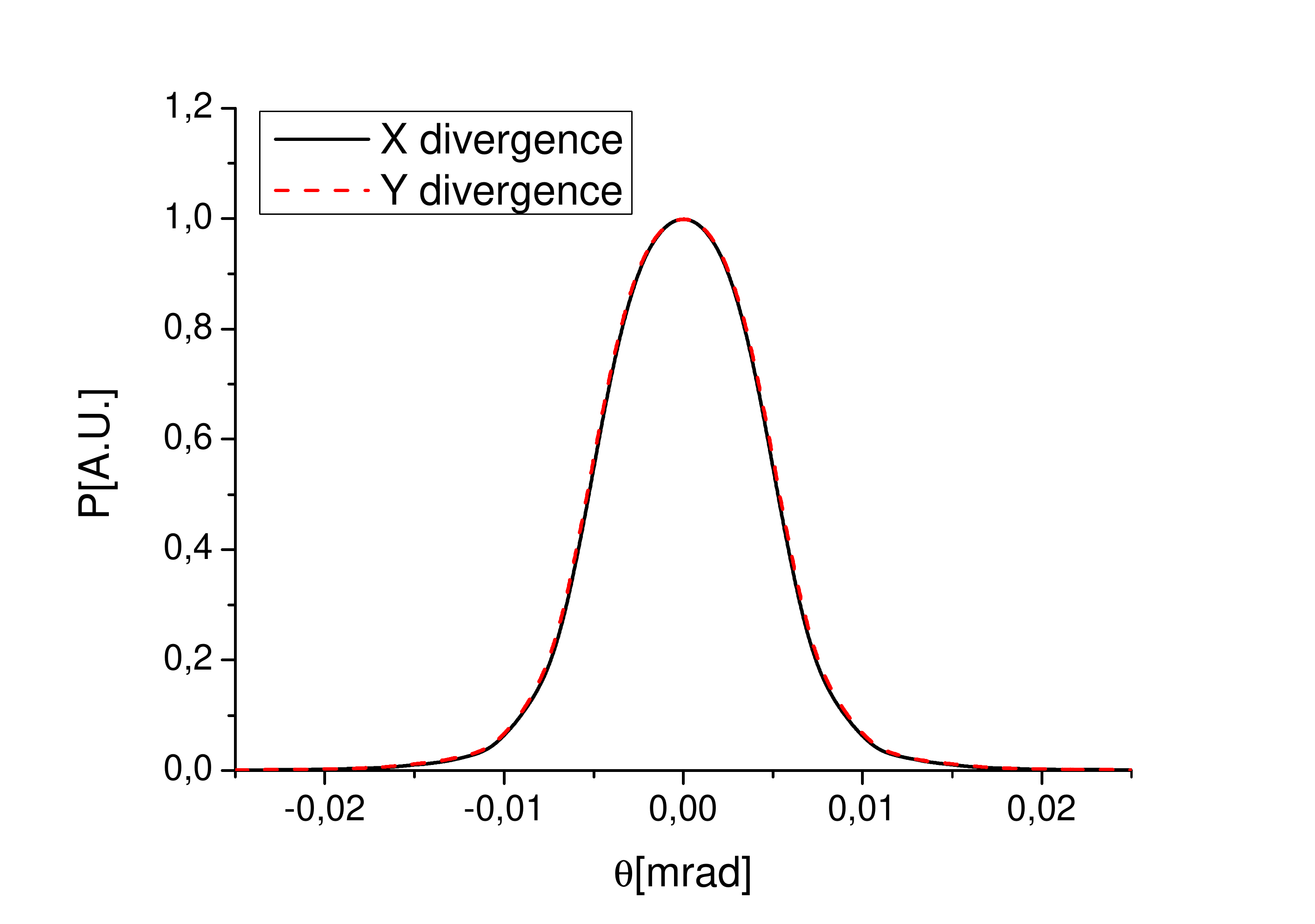}
\includegraphics[width=0.5\textwidth]{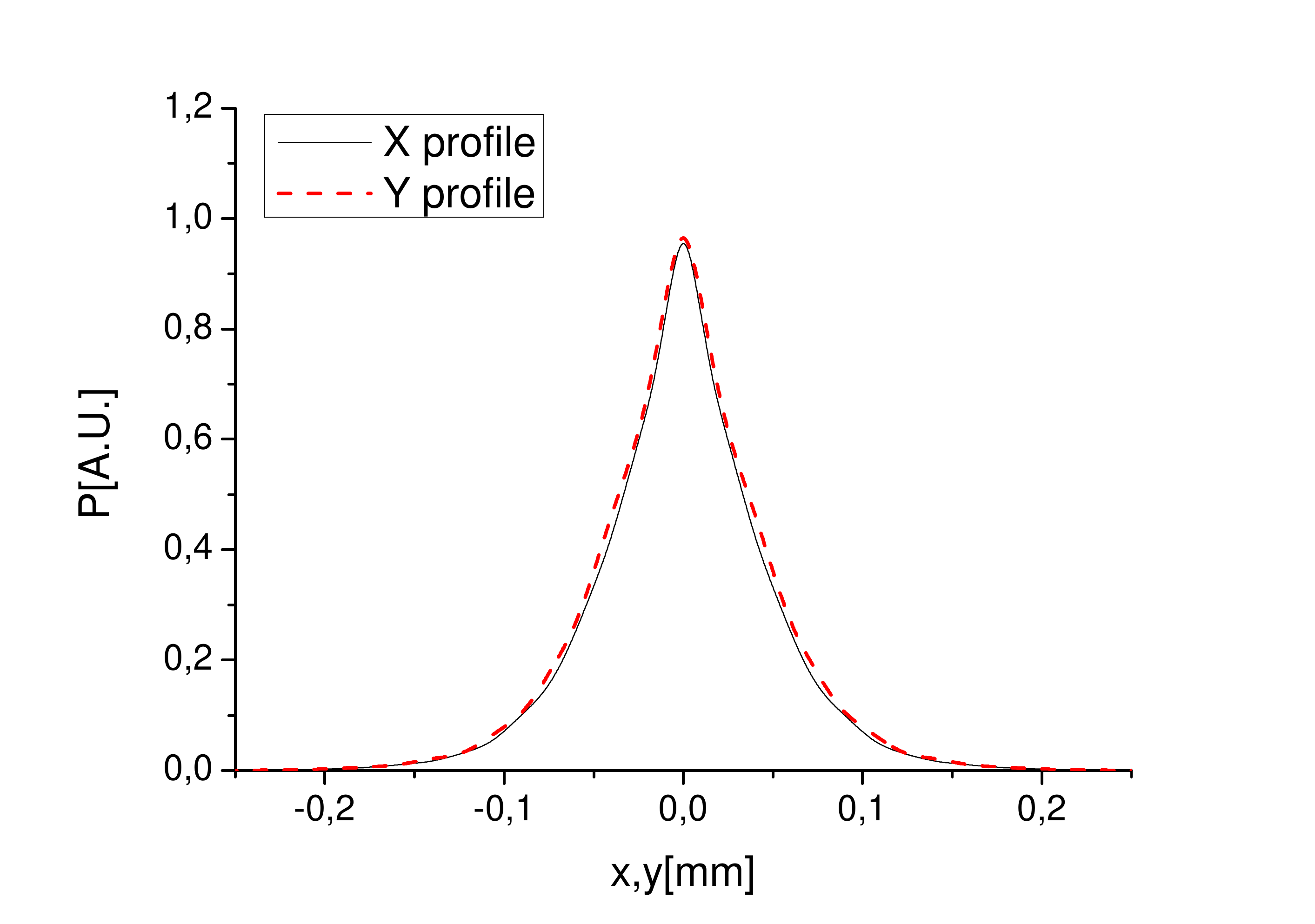}
\includegraphics[width=0.5\textwidth]{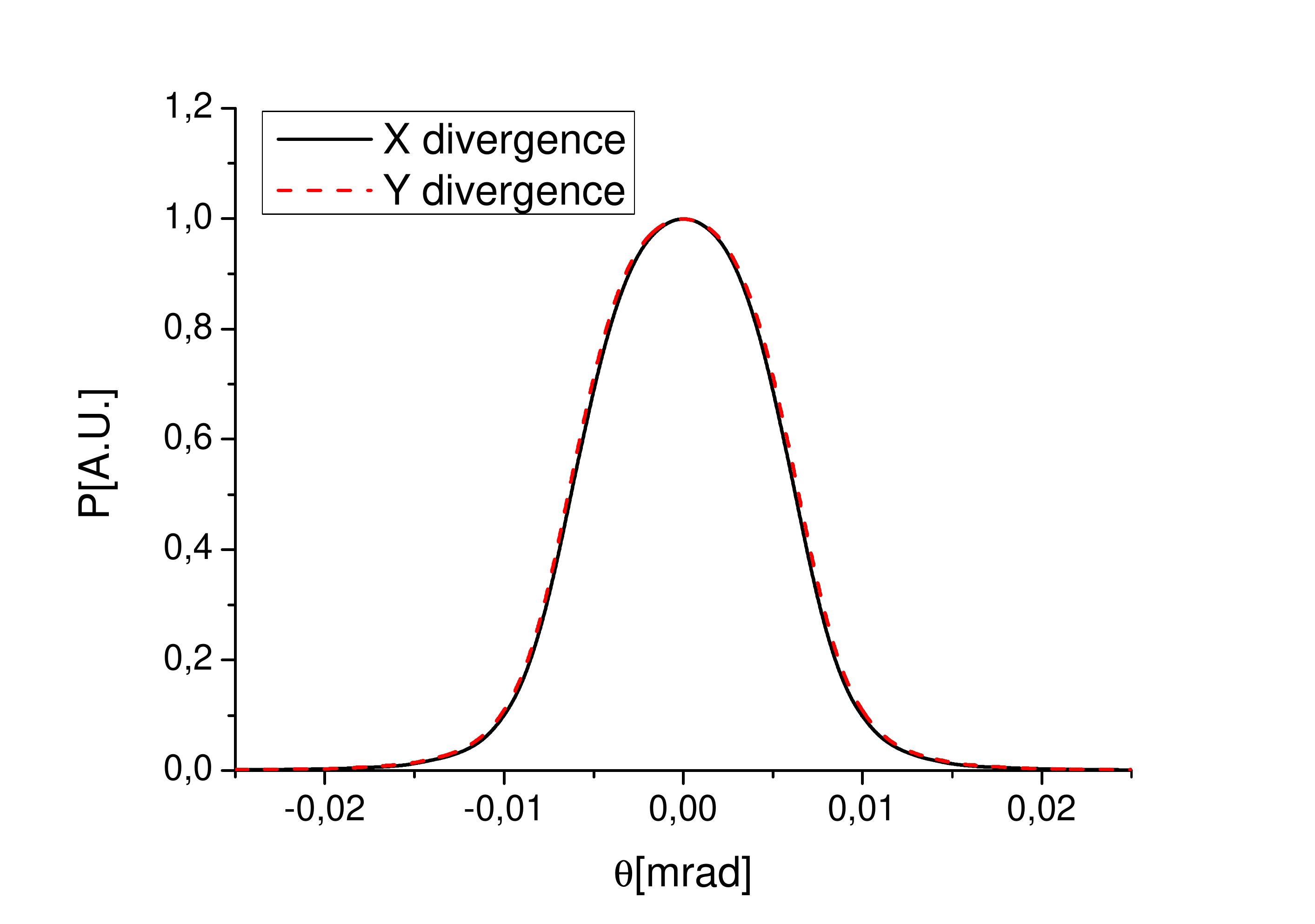}
\caption{Top row: Distribution of the radiation pulse energy per
unit surface and angular distribution at saturation. Bottom row:
Distribution of the radiation pulse energy per unit surface and
angular distribution at the exit of the setup.} \label{SpotTS}
\end{figure}
Finally, in Fig. \ref{SpotTS} we show a comparison of radiation beam
size and divergences for the SASE operation mode in saturation, and
for the tapered SASE mode. If one considers, as an example, the case
when pink light is delivered to the Small Quantum Systems (SQS)
station, the only optics encountered by the SASE3 beam transport
line is a pair of horizontal offset mirrors. This mirror system can
be adjusted between $6$ mrad and $20$ mrad incidence angle. After
the offset is introduced, the beam can be directly transported to
the SQS instrument \cite{SQST}. The offset mirrors are placed about
$300$ m behind the source point. Since one needs to minimize
diffraction from the optics aperture and to preserve the radiation
wavefront, any optical element should ideally have an aperture size
large enough to accept at least $4 \sigma$ times the beam size. In
our simulation study, the FWHM divergence of the FEL beam in the
tapered undulator case is of the order of $12~ \mu$rad at $2$ keV
(see Fig. \ref{SpotTS}). Therefore, the lateral aperture of the
radiation at the offset mirrors position turns out to be about $6$
mm (or $3.6$ mm FWHM) for our case of interest.

The lateral aperture of the radiation which is accepted by the
mirror is limited by the grazing angle and the length of the mirror.
More specifically, we can estimate the transverse clear aperture of
a grazing-incidence mirror as $L \theta$,  $L$ being the mirror
clear aperture, and $\theta$ being the grazing incidence angle. The
X-ray optics and transport group is planning to implement offset
mirrors with a clear aperture of $800$ mm  (see e.g. \cite{ODTR}).
With 9 mrad reflection angle we obtain a transverse clear aperture
of 7.2 mm, which is in principle enough to fulfill the $4\sigma$
requirement.

Finally, the transmission for the pair of offset mirrors for $9$
mrad incidence angle calculated, for B4C coating is about $90 \%$ at
a photon energy of 2 keV \cite{ODTR}.

\section{Cross-checking of Genesis simulations}

\begin{figure}
\begin{center}
\includegraphics[width=0.50\textwidth]{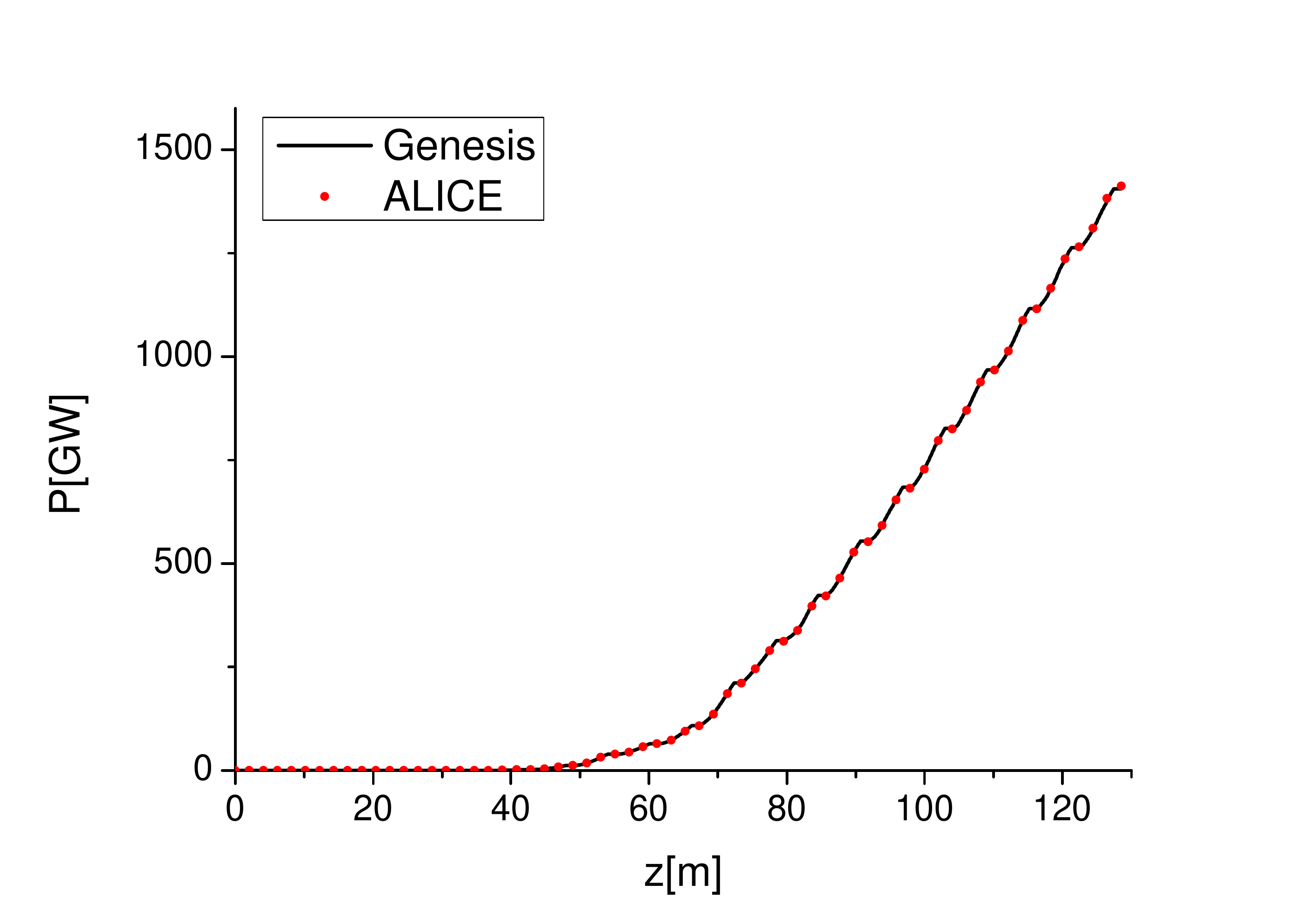}
\end{center}
\caption{Comparison between Genesis and ALICE predictions for the
output power as a function of the position inside the undulator in
the steady state regime. The FEL configuration considered here refer
to the SASE3 undulator line operating at 14 GeV. The photon energy
chosen for the comparison is 2 keV. Electron beam characteristics
refer to the longitudinal position, inside the 0.1 nC bunch,
corresponding to the maximum current value. The undulator magnetic
field file corresponds to the optimum for the SASE tapering regime
(see above in this section) } \label{GAcomp1}
\end{figure}
\begin{figure}
\begin{center}
\includegraphics[width=0.50\textwidth]{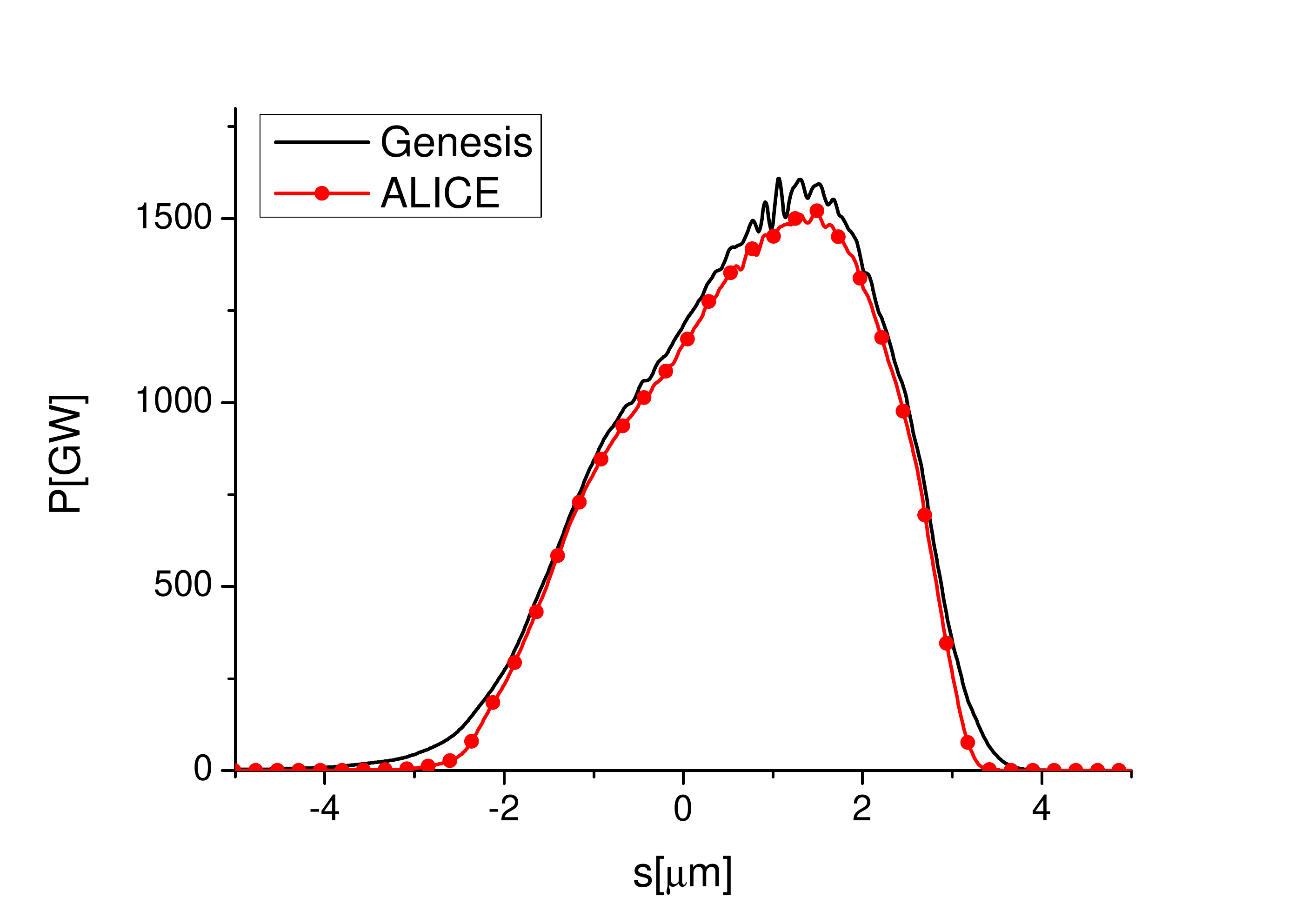}
\end{center}
\caption{Comparison between Genesis and ALICE predictions for the
output power as a function of the position inside the electron bunch
in time-dependent case for seeded FEL amplifier regime, including
tapering. The FEL configuration considered here refer to the SASE3
undulator line operating at 14 GeV. The photon energy chosen for the
comparison is 2 keV. Input electron beam and undulator magnetic
files are the same as in Fig. \ref{PowSpec}, bottom row.}
\label{GAcomp2}
\end{figure}

\begin{figure}
\includegraphics[width=0.50\textwidth]{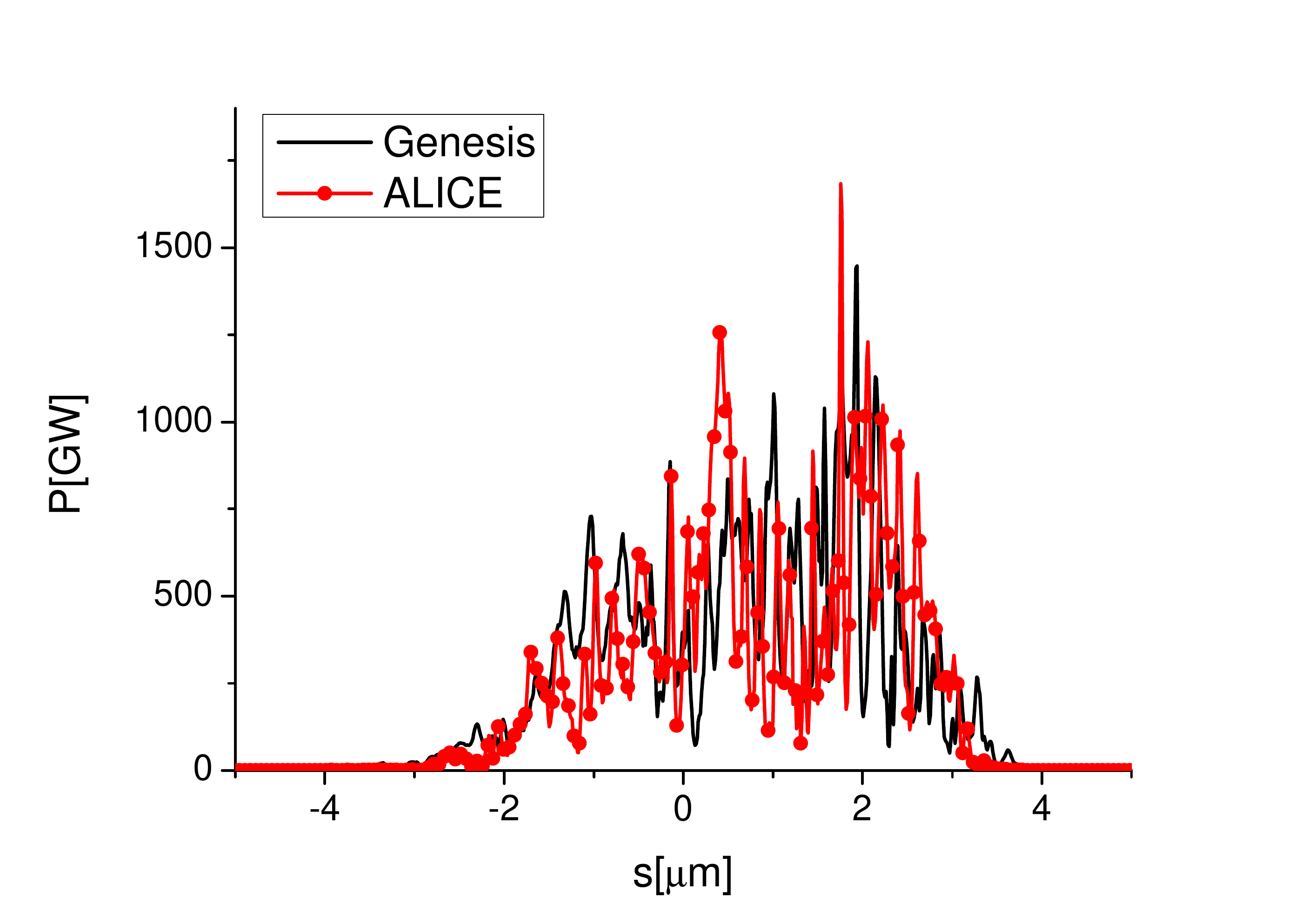}
\includegraphics[width=0.50\textwidth]{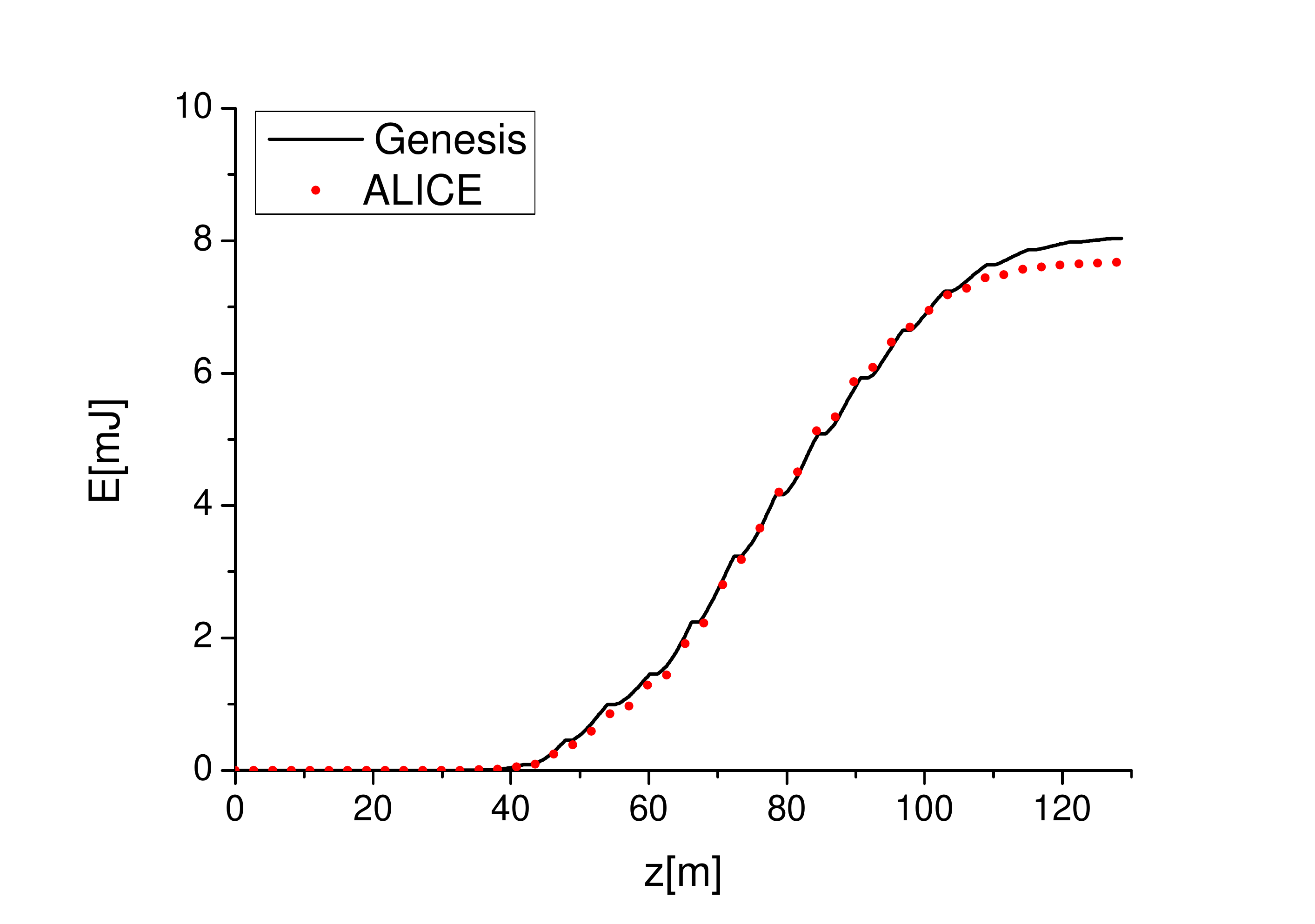}
\caption{Comparison between Genesis and ALICE predictions in the
time-dependent case for SASE regime, including tapering. The FEL
configuration considered here refer to the SASE3 undulator line
operating at 14 GeV. The photon energy chosen for the comparison is
2 keV. Left plot: output power for a typical single shot. Right
plot: energy per pulse. Since codes are not identical, SASE initial
conditions are not the same, and we can expect some difference
within results. Because of this reason, codes can be considered in
very good agreement. } \label{GAcomp3}
\end{figure}
While the code Genesis has had an undiscussed success to reproduce
results from LCLS experiments and has been thoroughly benchmarked,
next generation FEL codes like ALICE \cite{ALIC} recently began to
appear, which take advantage of more and more advanced algorithms.
In ALICE, the equations of motion for the particles are integrated
with a symplectic "leap-frog" scheme. The parabolic field equation
is solved with the help of an implicit Neumann finite difference
scheme, based on azimuthal expansion. Additionally, open boundary
conditions with the help of the Perfectly Matched Layer (PML) method
for parabolic equations were implemented. The code is parallelized
and allows one to use full three dimensional models for both the
electron beam and the radiation field.

In order to increase the confidence in our simulation results, we
cross-checked them with ALICE. In Fig. \ref{GAcomp1}, Fig.
\ref{GAcomp2} and Fig. \ref{GAcomp3} we show comparisons of
simulations obtained with these two codes, for undulator
configurations similar to those considered in this section. The
agreement in the steady state regime and in the time-dependent
seeded FEL amplifier regime is perfect. In the SASE regime the
agreement is reasonably good. Despite the simulation of deep
nonlinear SASE regime is the challenging problem for numerical
analysis and codes adopted different numerical methods, the pulse
energy differs by only $10 \%$. Differences in the output are well
within an rms of  shot-to-shot fluctuations (see Fig.
\ref{GAcomp3}). The SASE process is driven by the intrinsic shot
noise in the electron beam current, and a random generator is used
in each code to fill-in the initial particle distribution.
Therefore, differences in the SASE regime are ascribed to different
initial SASE signals.

\section{\label{sec:conc} Conclusions}

In this article we demonstrated that the output characteristics of
the European XFEL in SASE mode can be substantially improved,
without any additional hardware installation. At variance with
previous numerical studies for the European XFEL, in this work we
included energy chirp, wakefields and undulator tapering effects for
a segmented undulator with intersections, and we assumed realistic
distributions for the electron beam current, for the energy spread
and for the emittance in the horizontal and vertical direction along
the bunch. The importance of realistic models in XFEL simulations is
amply demonstrated in this paper. In order to take into
consideration wakefield effects and to examine the impact of
undulator tapering on radiation properties, electron bunches are
treated on the basis of start-to-end simulations and the undulator
includes intersections with phase shifters. In order to obtain
optimal performance, both undulator gap and phase shifters had to be
tuned. It has been shown that the taper provides an additional
factor of ten increase in spectral density and output power (up to
the TW-level) for a baseline electron beam parameter set.  We
anticipate that, mainly by increasing the peak current of the
driving electron beam beyond the baseline value of $5$ kA it will be
possible to achieve a further increase of x-ray power beyond the
TW-level. The European XFEL uses a superconducting L-band linac to
accelerate electron beams, which can be compressed up to extremely
high peak currents of about 50 kA, with still a reasonable electron
bunch quality \cite{IGO2}. The maximal electron beam energy at the
European XFEL is 17.5 GeV. It is worthwhile to mention that 50 kA
peak current at 20 GeV beam energy corresponds to 1 PW peak power of
electron beam. In \cite{10TW} we demonstrated that with a PW power
level of driving electron beam and a long enough, high K-value
undulator it would possible to achieve a 10 TW-level X-ray beam at
the European XFEL without additional hardware. For a long time, such
unique feature will be available only at the European XFEL.

\section{Acknowledgements}

We are grateful to Massimo Altarelli, Reinhard Brinkmann, Henry
Chapman, Janos Hajdu, Viktor Lamzin, Serguei Molodtsov and Edgar
Weckert for their support and their interest during the compilation
of this work.

\end{document}